\documentclass[10pt]{iopart}
\usepackage{iopams}
\usepackage{graphicx}
\usepackage{hyperref}
\newcommand{\R}{\mathbb{R}}
\newcommand{\nn}{N_{\rm N}}
\newcommand{\nl}{N_{\rm L}}
\begin{document}
	
	\title[Reflectivity profile inversion by ANNs]{Towards Reflectivity profile inversion through Artificial Neural Networks}
	
	\author{Juan Manuel Carmona Loaiza\textsuperscript{1, *} and Zamaan Raza\textsuperscript{1}}
	
	\address{\textsuperscript{1}J\"ulich Centre for Neutron Science (JCNS) at Heinz Maier-Leibnitz Zentrum (MLZ).
		Lichtenbergstraße 1, 85748 Garching, Germany.
	}
	\ead{j.carmona.loaiza@fz-juelich.de}
	\vspace{10pt}
	\begin{indented}
		\item[]22 December 2020
	\end{indented}
	
	\begin{abstract}
		The goal of Specular Neutron and X-ray Reflectometry is to infer materials Scattering Length Density (SLD) profiles from experimental reflectivity curves. This paper focuses on investigating an original approach to the ill-posed non-invertible problem which involves the use of Artificial Neural Networks (ANN). In particular, the numerical experiments described here deal with large data sets of simulated reflectivity curves and SLD profiles, and aim to assess the applicability of Data Science and Machine Learning technology to the analysis of data generated at neutron scattering large scale facilities. It is demonstrated that, under certain circumstances, properly trained Deep Neural Networks are capable of correctly recovering plausible SLD profiles when presented with never-seen-before simulated reflectivity curves. When the necessary conditions are met, a proper implementation of the described approach would offer two main advantages over traditional fitting methods when dealing with real experiments, namely, 1. sample physical models are described under a new paradigm: detailed layer-by-layer descriptions (SLDs, thicknesses, roughnesses) are replaced by parameter free curves $\rho(z)$, allowing a-priori assumptions to be fed in terms of the sample family to which a given sample belongs (e.g. "thin film", "lamellar structure", etc.) 2. the time-to-solution is shrunk by orders of magnitude, enabling faster batch analyses for large datasets.
	\end{abstract}
	
	\vspace{2pc}
	\submitto{\MLST}
	%\maketitle
	\ioptwocol

	\section{Introduction}
	
	Neutron and X-ray Specular Reflectometry are established experimental techniques whose aim is to investigate interfacial structures at the sub-nanometer scale through the measurement and analysis of reflectivity curves \cite{Parratt1954, Penfold1990, Tanner2018}.
	
	In a typical specular reflectometry experiment, a collimated Neutron or X-ray beam of wavelength $\lambda$ impinges on the surface of a flat sample at an incident angle $\theta$. The incident angle is varied and the specular reflectivity is measured as the ratio between the reflected and the incident beam intensities, $R(\theta) = I_R(\theta)/I_0(\theta)$.
	
	Theoretically, in the absence of significant non-specular scattering from in-plane variations of the SLD, neutron specular reflectivity is accurately described by a one-dimensional Schr\"odinger wave equation\footnote{Throughout this work, derivations and discussions focus mainly on neutron reflectometry, however, the same derivations and approach apply to the X-ray case with little or no change.},
	
	\begin{equation}\label{eq: schroedinger}
	-\frac{\partial^2\psi(k_{0z}, z)}{\partial z^2} + 
	4 \pi \, \rho(z) \psi(k_{0z}, z) = 
	k_{0z}^2 \, \psi(k_{0z}, z),
	\end{equation}
	
	\noindent where $\psi$ is the wave function, $\rho$ is the SLD profile of a given sample, $k_{0z}$ is the wave vector $z$ component and $z$ is the depth inside the sample, perpendicular to the sample interfaces (For more details see e.g. \cite{Majkrzak2003} and references therein).
	
	In terms of the solution to equation \ref{eq: schroedinger}, the amplitude of the
	reflected wave can be represented by the integtral
	
	\begin{equation}\label{eq: r_of_q}
	r(Q) = \frac{4\pi}{iQ} \int_0^L{\psi(k_{0z}, z) \, \rho(z) \, {\rm e}^{ik_{0z}\,z} {\rm d}z},
	\end{equation}
	
	\noindent where $r(Q)$ is the complex-valued reflection amplitude as function of the wave vector 
	transfer perpendicular to the surface, $Q = 2 \, k_{0z} = 4 \pi \sin(\theta) / \lambda$, and $L$ is the thickness of the SLD profile.
	However, the only quantity accessible to measurements is the
	reflectivity, which can be expressed in terms of the amplitude as $R(Q) \equiv r^*r$.
	
	\subsection{The phase problem and fitting}\label{sec: phase_problem}
	
\begin{figure}
		\centering
		\includegraphics[width=0.5\textwidth]{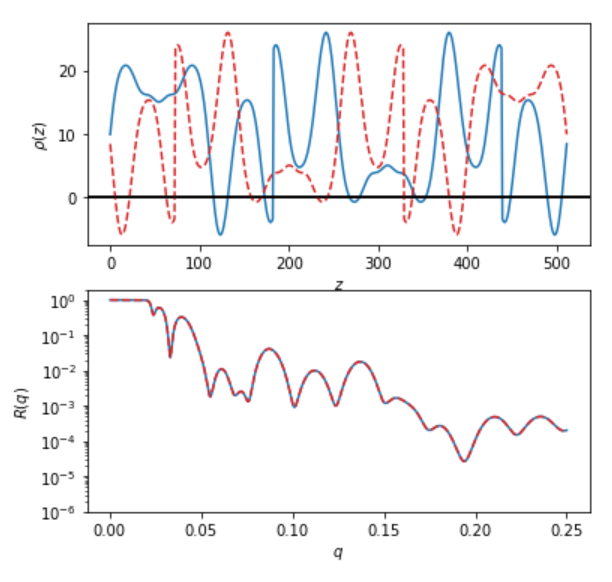}
		\caption{Two different anti-periodic SLD profiles, one the mid-point reflection of the other, produce the exact same reflectivity curve.}
		\label{fig: PhaseProblem}
	\end{figure}

	The measured reflectivity, $R(Q)$, does not carry any information regarding the phase, making the inference of an SLD profile from a reflectivity curve a non-invertible inverse problem: at a theoretical level, there are families of SLD profiles which produce exactly the same reflectivity curve. In particular, this applies to any anti-periodic SLD profile that is reflected at the mid point (See Figure \ref{fig: PhaseProblem}). To an experimenter measuring reflectivities, $R(Q) = r^*r$, both SLD profiles are indistinguishable.
	
	The degeneracy causing different SLD profiles to produce identical reflectivity curves is known as the phase problem, and is accentuated by the truncation of the reflectivity data at a maximum value of $Q$ and the statistical uncertainty associated with noisy data points (e.g. See \cite{Majkrzak2003unambiguous}).
	
	Typically, the data obtained from Reflectometry experiments is analyzed in terms of physical models, trusting their ability to reproduce measured experimental reflectivity curves. Using specialized software (e.g.\cite{BornAgain2020, Refnx2019}), an iterative process of parameter optimization is thus established in which, at each iteration, 1. certain parameters of the physical model are set, 2. theoretical reflectivity curves are calculated and 3. a comparison is made between the theoretical and experimental reflectivity curves. The latter comparison is quantified by a Figure of Merit (FOM) and the goal of the iterative process is reached when the FOM reaches its minimum value, namely, when the experimental and the theoretical reflectivity curves are as close as allowed by the physical model and the experimental resolution. Such an iterative \textit{fitting} process is far from immediate, requiring experimenters to try out several FOMs, several minimization algorithms, several sets of model parameters and even several physical models. 
	%[JUAN: BornAgain paper, some references from Alexandros 2019: \cite{Nelson06, Gerelli16a, Gerelli06b}]
	
	%Success in recovering the correct SLD profile depends on the available a-priori information and on the validity of the initially built model. However, to reach an objective final solution, the concurrent fitting of multiple-contrast (e.g.\cite{Fragnetto95, Braun17, Wacklin10, Bradley18}) reflectometry experiments is often necessary.
	
	In recent years, BornAgain \cite{BornAgain2020} --a well established code for simulating and fitting neutron and X-ray grazing-incidence small-angle scattering (GISAS) data, has started to support fitting and simulation capabilities for reflectometry data as well. The present work aims at exploring the possibilities that Machine Learning has to offer towards the development of reflectivity data-driven software.

	\section{The expressive power of artificial neural networks}
	
	Deep neural networks can be thought of as compositions of multiple simple functions (called layers) that can approximate rather complicated functions. In fact, the celebrated universal approximation theorem states that depth-2 networks with suitable activation functions can approximate any continuous function on a compact domain to any desired accuracy \cite{Cybenko1989, Funahashi1989, Hornik1989, Barron1994}. However, the size of such a neural network could be exponential in the input dimension, which means that the depth-2 network may have a very large width \cite{Cohen2016, Eldan2016, Telgarsky2016}. In fact, part of the recent renaissance in Artificial Neural Networks (ANN), lies not on enabling wider networks to be trained, but on the empirical observation that \textit{deep neural networks} tend to achieve greater expressive power per parameter than their shallow counterparts.
	
	It has been shown that any Lebesgue-integrable function from $\R^N \rightarrow \R$ can be approximated by a fully-connected \texttt{ReLU} deep neural network of width $N+4$ to arbitrary accuracy with respect to $L_1$ distance and, except for a negligible set, all functions from $\R^N$ to $\R$ cannot be approximated by any \texttt{ReLU} network whose width is no more than $N$ \cite{Lu2017}. Hanin B. (2019, \cite{Hanin2019}) show that any continuous function $f: [0,1]^{d_{in}} \rightarrow \R^{d_{out}}$ can be approximated by a net of width $d_{in}+d_{out}$, obtaining also quantitative depth estimates for such an approximation in terms of the modulus of continuity of $f$. At the same time, they claim that there are no conclusive results regarding the depth such a network should have, and, even in the case that a precise ANN architecture to achieve a given precision can be defined, nothing can be said regarding the success of the training process.
	
	In contrast to these apparent theoretical drawbacks, one of the first architectures that is taught when studying ANNs --a single hidden layer of width $128$, is able to classify $28 \times 28$ pixel images (i.e. points in $\R^{784}$) into 10 discrete categories, passing in the process through a mapping into the real unit interval (e.g. \texttt{TensorFlow} tutorials \cite{tensorflow2015-tutorial}). These achievements, in apparent contradiction with the theoretical results described above, are possible because of the fact that the images under classification do not sample the whole $\R^{784}$ but are drawn from only a very limited subspace of it. In fact, when the trained network is presented with images that do not belong to that subspace, the ANN fails --it may even classify apparently random noise as some of the 10 digits with almost 100\% certainty.
	
	In the present work, focus is made on training simple and small ANNs targeted to specialized kinds of SLD profiles, as opposed to larger and general-purpose neural networks, as any attempt to teach a single ANN a general pseudo-inverse function, for the time being, is almost certainly doomed to fail.
	
	\section{Related work}
	
	Non invertible inverse problems are not unique to X-ray and Neutron reflectometry, and several other scientific communities have already started to investigate the usefulness of Artificial Neural Networks for tackling them, showing astonishing performance for applications like low-dose computed tomography or various sparse data problems. While there are few theoretical results, some well-posedness results and quantitative error estimates have been found for some problems \cite{Li2020nett}. For instance, in Electrical Impedance Tomography (EIT), which represents the typical nonlinear ill-posed problem, the electrical properties of tissues are determined by injecting a small amount of current and measuring the resulting electric potential, which must be transformed into a tomographic image by some reconstructing algorithm. Many artificial intelligence approaches to tackle EIT have been taken in the past few years (e.g. \cite{Khan2019} and references therein) with outstanding results.
	
	In the realm of X-ray reflectivity, recent work shows that properly trained ANNs with simple fully connected architectures can be used to characterize thin film properties (thickness, roughness and density) from XRR data within milliseconds and minimal a priori knowledge. Their results differ from traditional least mean squares fitting by less than 20\% \cite{Greco2019}. Such an approach could benefit the study of the growth behavior of thin films.
	
	Current software packages that are available to infer physical models from reflectivity curves allow users to define SLD multi-layer models and, after some fitting process, choose the parameter combination that best reproduce the experimental data; all of this with varying degrees of interactivity either through an application programming interface (API) or a graphical user interface (GUI). Examples of parameters to fit are layer thicknesses, roughnesses between layers and layer SLD values. In contrast, the physical models used in the present work to represent SLD profiles do not make use of the multi-layer abstraction. The models used in the present work deal with quasi-continuous SLD profiles that have an SLD value defined at each point and varies continuously throughout the sample without the need of introducing extra abstractions like interfacial roughness.
	
	\section{Application to reflectivity profile inversion.}
	
	\subsection{Scaling of the problem}

	The only physical quantities involved in the calculation of a reflectivity curve, assuming a perfect instrument able to measure up to $Q \rightarrow \infty$ and an SLD profile extending up to $z \rightarrow \infty$, are the wave transfer vector $Q$ and the SLD profile $\rho(z)$. These quantities can be further reduced by using dimensional analysis. In fact, the number of dimensionless groups that define the problem, which equals the total number of physical quantities ($Q$ and $\rho$) minus the fundamental dimensions (length), is only one ( = $2-1$). By choosing an arbitrary SLD scale $\rho_0$ and defining the dimensionless parameter $p = \rho / \rho_0$, equation (\ref{eq: r_of_q}) can be re-casted in the following form: 
	
	\begin{equation}\label{eq: r_of_q_dimless}
	r_0(Q) = \sqrt{\rho_0} \times  \frac{4\pi}{i Q} \int_0^\infty{\psi(\eta, \xi) \, p(\xi) \, {\rm e}^{i\eta\,\xi} {\rm d}\xi},
	\end{equation}
	
	\noindent where $\xi = z \sqrt{\rho_0}$, $\eta = k_{0z} / \sqrt{\rho_0}$, and $p(\xi) = \rho(z) / \rho_0$ are the dimensionless depth, wave vector and SLD profile respectively.
	Thus, to solve for a different SLD scale, $\rho_*$, it is enough to solve for $r_0(Q)$ and rescale afterwards by $\sqrt{\rho_*/\rho_0}$, i.e. 
	\begin{equation}\label{eq: rescaled}
	r_*(Q) = \sqrt{\frac{\rho_*}{\rho_0}} \times r_0(Q) .
	\end{equation}
	In the following, the SLD scale of the problem is chosen to be that of the substrate, $\rho_{0} = \rho_{subs}$, i.e., $p_{subs} \equiv 1$.
	
	\subsection{Data simulation}\label{sec: data simulation}
	
	The phase problem implies that a single input (e.g. a reflectivity curve) may be consistent with two or more different outputs (e.g. SLD profiles). If one were to train an ANN to find a pseudo-inverse transformation, using data containing different output targets corresponding to the same input (different branches), the training process would not be successful, as different branches would cause the weights of the ANN to drift in inconsistent directions. In order to avoid such a situation, it must be ensured that either $*1$.- the solution space  has no branches, or $*2$.- the training targets lie all in the same branch of the solution space. For the last scenario, it must also be kept in mind that ANNs trained in such a way will only be useful as long as the expected solutions are consistent with the branch to which the training targets belong. Looking to fulfill $*1$, a set of artificial SLD profiles is generated which could offer a 1-1 correspondence to their associated reflectivity curves, e.g. SLD profiles odd with respect to the middle of the depth\footnote{Note that it has not been rigorously shown that a 1-1 correspondence exist. Several profiles from the family could still produce the same reflectivity profile.}. To try to fulfill $*2$ to some extent, two sets of SLD profiles are generated, each set corresponding to a physically relevant typology of samples, namely, single films and lamellar structures.
	
	All simulated SLD profiles used as training targets in this work have an overall thickness $L = 512 \rm\AA$, and are sampled by 512 equally spaced points within the semi-closed interval $z = (0, L]$, between two semi-infinite fronting and backing mediums of constant SLD, $\rho_{- \infty} = 0$ and $\rho_{+ \infty} = 10^{-6} \rm\AA^{-2}$ respectively. Each SLD profile was thus modeled as a \texttt{Multilayer} composed of 512 slices, $1\,\rm\AA$ thick each, with no interfacial roughness. In this way, smooth SLD profiles were mimicked by quasi-continuous small variations of the SLD between contiguous layers throughout the whole interfacial structure. Such SLD profiles are then used to simulate the corresponding reflectivity curves for which the wave vector transfer is limited to an interval $0 \leq Q \leq Q_{\rm max} = 0.25$, sampled by 129 equally spaced points. Imperfections in the reflectivity curves are only characterized by a background of $10^{-6}$ and a $Q$-resolution of 5\%.
	
	In order to bring the so-called \textit{features} (in the case at hand, the reflectivity curves) into a small dynamic range, the average reflectivity curve needs to be subtracted from all reflectivity curves in the set. Thus, the new set of curves is a zero-mean set of curves. After that, each of the resulting zero-mean curves is rescaled by dividing it by the standard-deviation curve. Finally, the obtained zero-mean, unit-standard-deviation set of curves is ready to be used for training. In the following, this set of curves are referred to as the \textit{training features}. In contrast to the reflectivity curves, the SLD profiles, i.e. the \textit{training targets}, are left unchanged.
	
\begin{figure*}
		\includegraphics[width=0.32\textwidth]{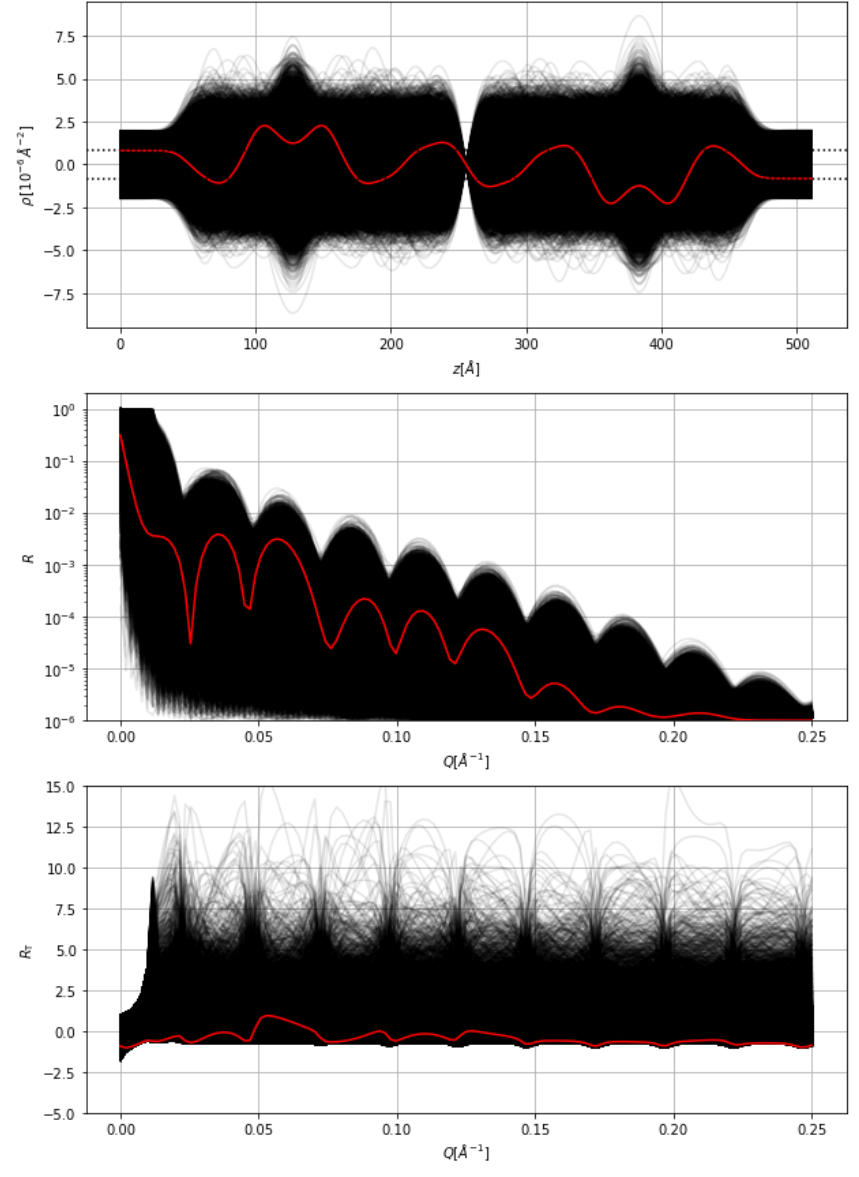}
		\includegraphics[width=0.32\textwidth]{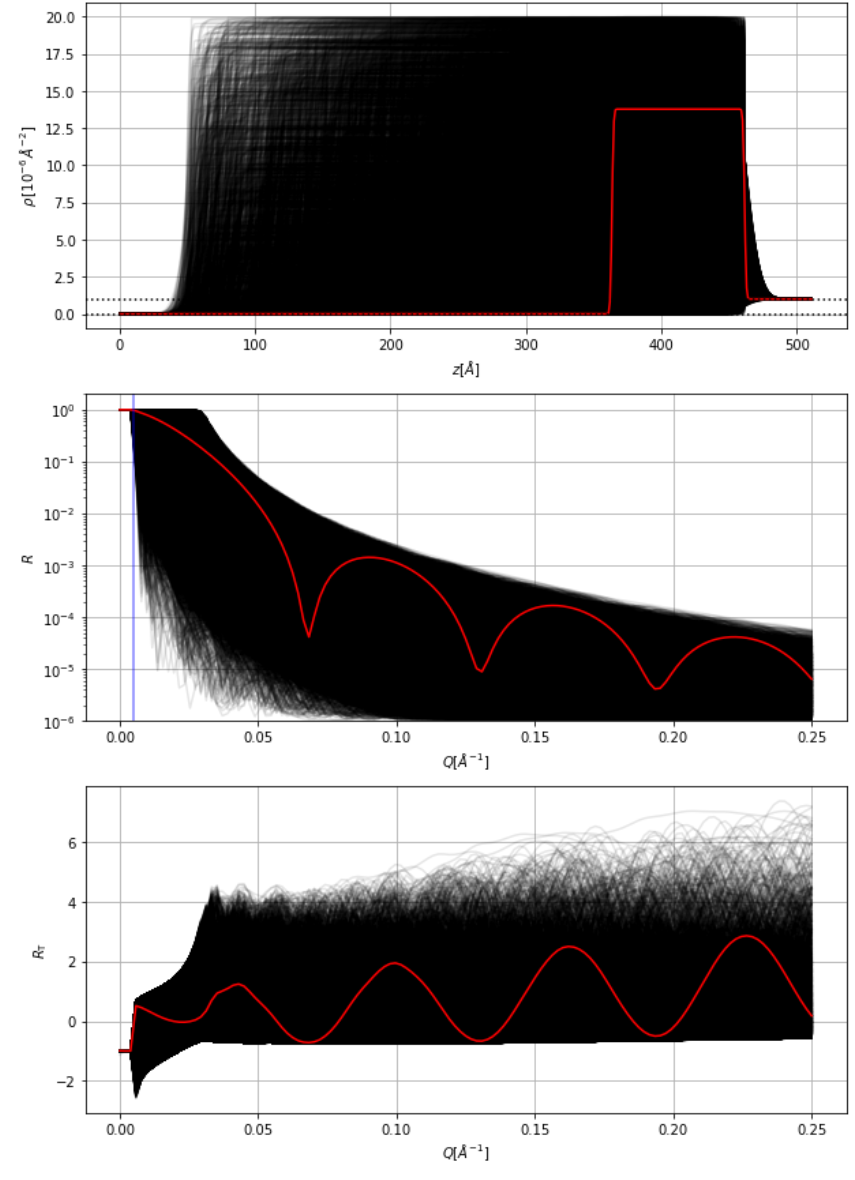}
		\includegraphics[width=0.32\textwidth]{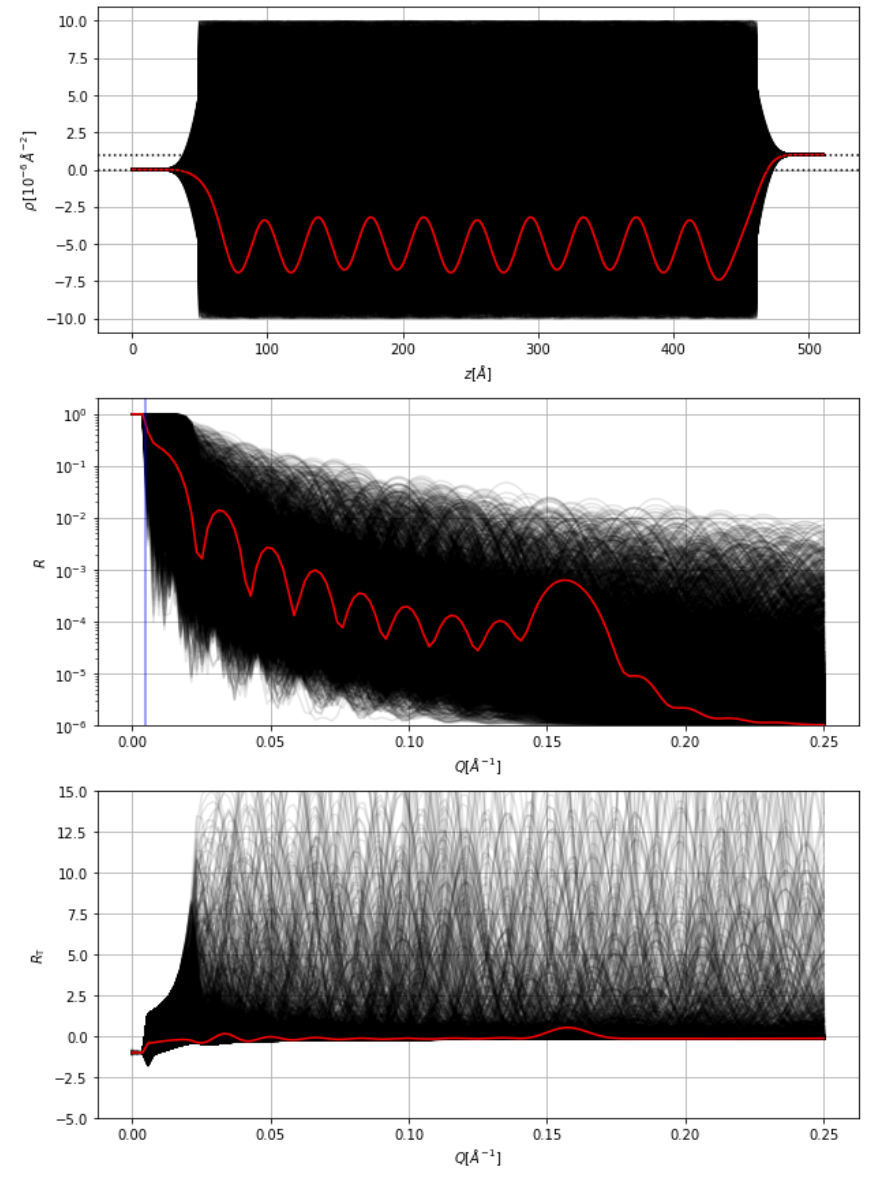}
		\caption{In the \textbf{top row}, three different families of SLD profiles are shown: Odd functions, single films and lamellar structures; the \textbf{middle row} shows the simulated reflectivity curves obtained from the SLDs in the top panels; The \textbf{bottom row} show the rescaling of the reflectivity curves which have zero mean and unit standard deviation. A single random curve is highlighted in red to give a qualitative impression of the features defining each family of curves.}
		\label{fig: data}
	\end{figure*}

	Figure \ref{fig: data} gives an overview of the three different data sets used and their preprocessing, and a more detailed discussion of each is carried out in section \ref{sec: results}.
	
	\subsection{Network architecture}\label{sec: architecture}
	
	The network architecture is given to some extent by the problem constraints: The number of input neurons must be the same as the input reflectivity curve lengths, which is chosen to be $N_{\rm in} = 129$. The number of output neurons must be the same as the expected SLD profile resolution points, which are $N_{\rm out} = 512$. 
	
	Hidden layers are chosen to be fully connected layers, a choice dictated by the problem constraints as well. In fact, the choice of such kind of layers, as opposed to the so-called \textit{convolutional} layers, powering state-of-the-art image recognition and segmentation algorithms is well grounded in a simple observation: while convolutional layers are powerful in detecting shapes and measuring their relative positions with respect to one another, independent of their absolute positions within the scene or image, in the case at hand, it is actually needed the opposite: the absolute position of all intensity data points along the $q$ line is all that matters to define a reflectivity curve.
	
	To reproduce non-linearities, \texttt{ReLU} functions are used after each hidden layer, except the one before the output layer and, in order to prevent \textit{overfitting}, a single dropout layer with a rate of 0.5 is added as a regularizer before the output layer.
	
	The architecture so far is dictated to great extent by the problem. However, there are no definite rules to select neither a concrete number of hidden layers nor the number of neurons present in each hidden layer. In order to make these concrete choices, preliminary experiments are carried out for different network depths and widths (i.e. number of hidden layers, $N_{\rm L}$, and number of neurons per hidden layer, $N_{\rm N}$, respectively). The architectures are arbitrarily restricted to hidden layers possessing all the same number of neurons, with $N_{\rm L} = 0, 1, 2, 4, 8, 10$; $N_{\rm N} = \{0.25, 0.5, 1, 2, 4\} \times N_{\rm out}$\footnote{Throughout the rest of the manuscript, $\nn$ refers to the multiplicative factor of $N_{\rm out}$, and not to the actual number of neurons per hidden layer itself.}. Each of these prototype networks was then trained and tested using 100K artificial and non-physical SLD profiles of odd-function shape (left panel of Figure \ref{fig: data}). The performance of each ANN is tested by feeding them 5K never-seen-before reflectivity curves and calculating the Mean Absolute Error, 
	
	\begin{equation}\label{eq: mean_absolute_error}
	{\rm MAE}_j = {\rm MAE}(\rho_j, \tilde{\rho_j}) = \frac{1}{\rm N_{RES}}\sum_k{\rho_{jk} - \tilde{\rho}_{jk}},
	\end{equation}
	
	\noindent between the ANN predicted SLD profile and the known SLD target profile.
	
\begin{figure*}
	\includegraphics[width=0.33\textwidth]{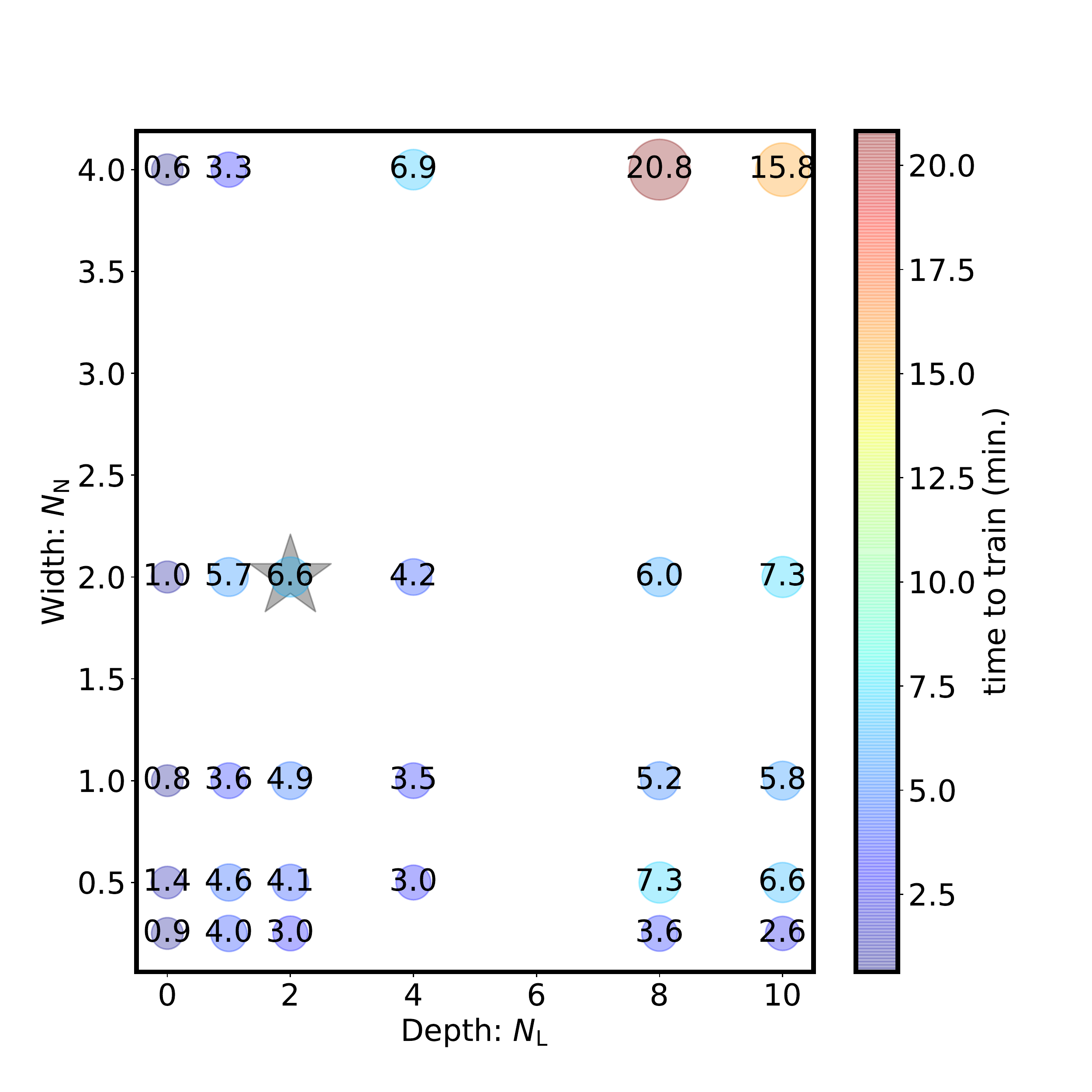}
	\includegraphics[width=0.33\textwidth]{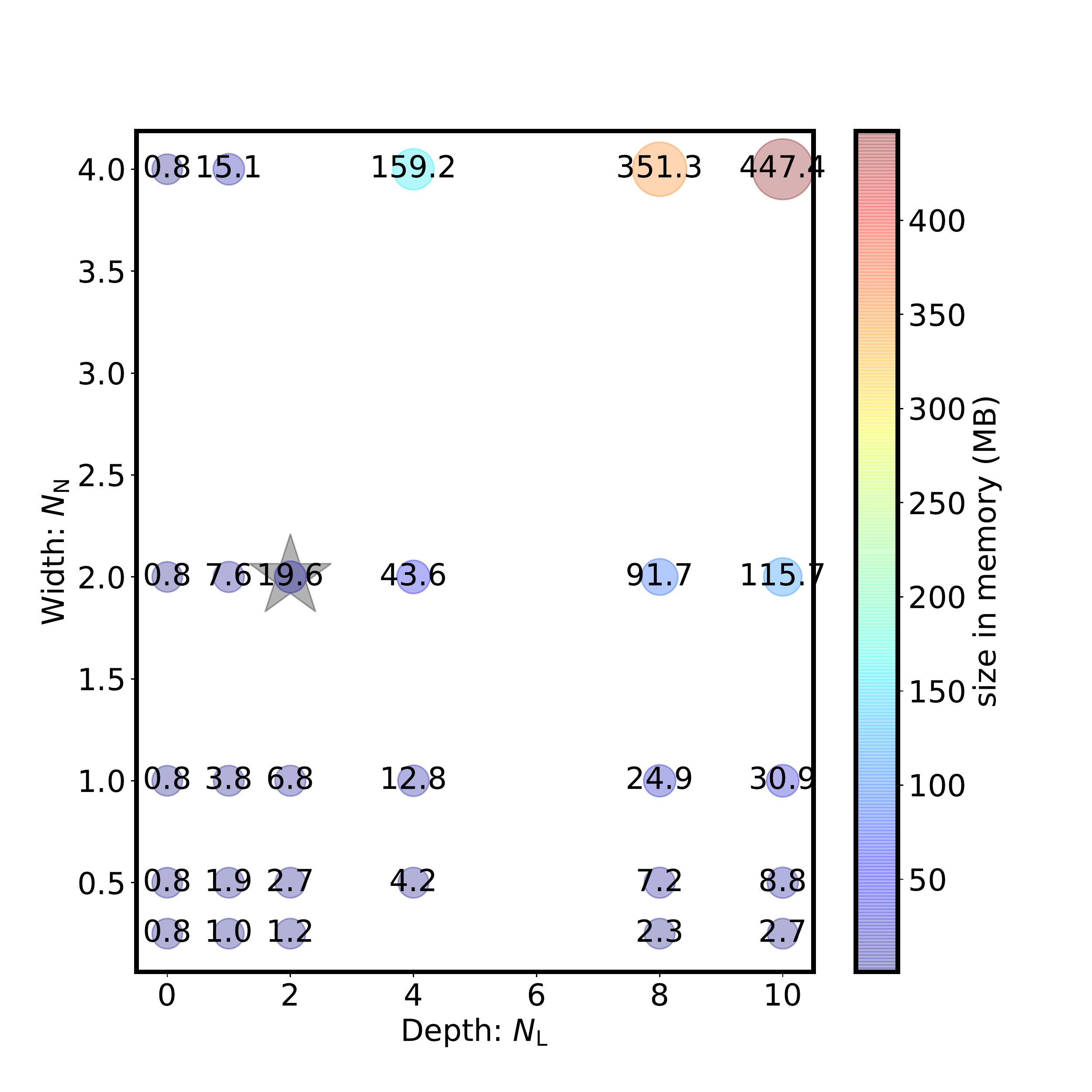}
	\includegraphics[width=0.33\textwidth]{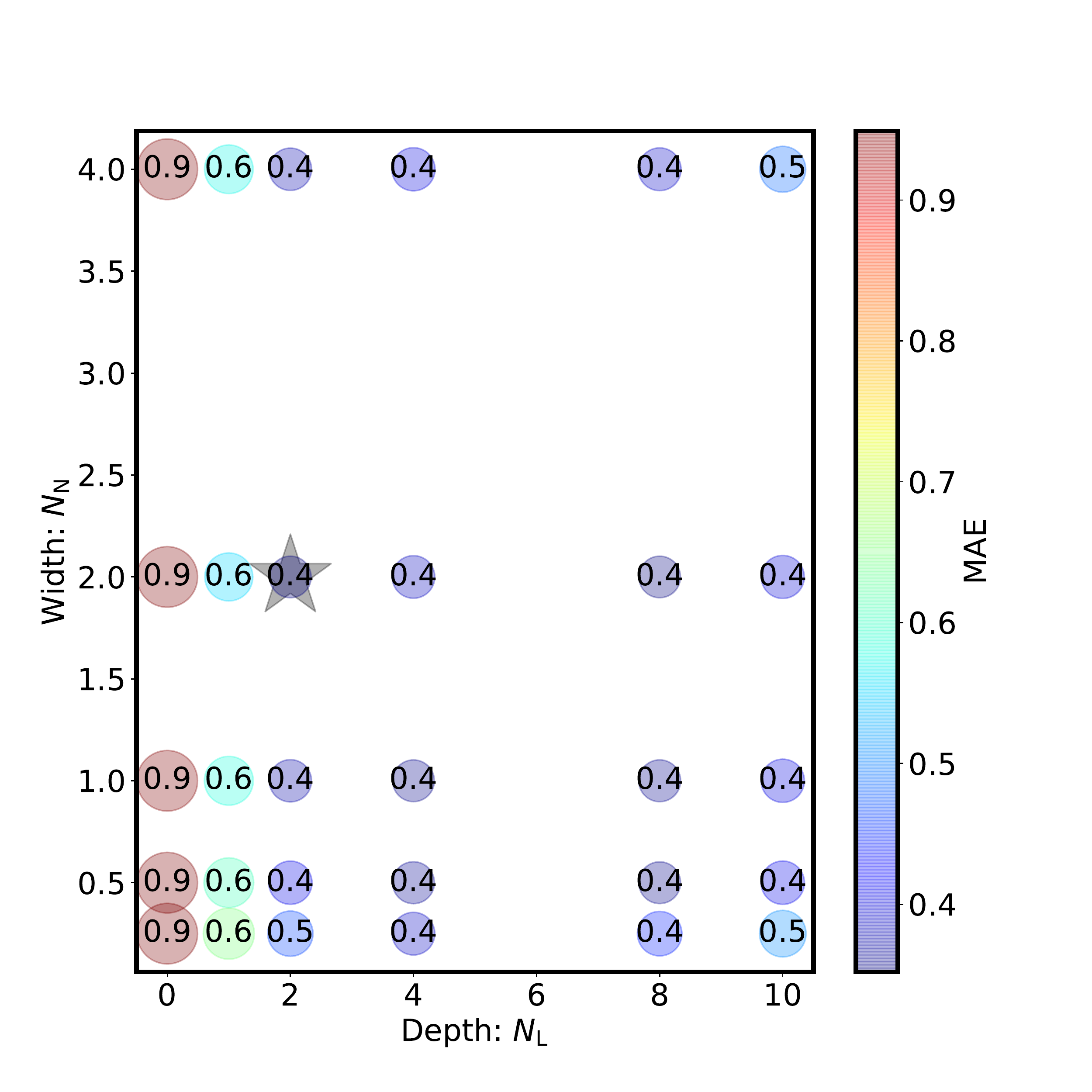}
	\caption{Several Network architectures were built and tested using 100K artificially generated SLD profiles and reflectivity curves.
	 On the \textbf{left panel}, the times each architecture takes to complete training are shown; on the \textbf{middle panel}, their disk-space requirements, and, on the the \textbf{right panel}, the MAE incurred between predicted SLD profiles and the true targets present in the test set. Ideally,a good ANN is accurate, trains fast and requires small amounts of space on disk. Each architecture tested is described by its number of hidden layers, $\nl$ and the number of neurons in each hidden layer $\nn$.}
	\label{fig: times_size_err_wrt_architecture}
	\end{figure*}

	The times-to-train, the memory space required and the MAE incurred by each of the architectures tested, are shown in Figure \ref{fig: times_size_err_wrt_architecture}. From this analysis, the best architecture was selected, $N_{\rm L} = 2$ and $N_{\rm N} = 2$, and used for the numerical experiments described in the remaining part of the present work.
	
%
%	; an architecture which is displayed by the \texttt{keras} method \texttt{Model.Summary()}
%	in the following way\footnote{Note that the first \textit{reshape} layer is only needed to reshape the input array from (1,129) to}:
%
%	\begin{minipage}{0.2\textwidth}
%		\vspace{0.3cm}
%		\fontsize{8}{10}
%		\begin{verbatim}
%		_________________________________________________
%		Model: "sequential"
%		_________________________________________________
%		Layer (type)          Output Shape     Param #   
%		=================================================
%		reshape (Reshape)     (None, 129)      0         
%		_________________________________________________
%		dense (Dense)         (None, 1024)     266240    
%		_________________________________________________
%		dense_1 (Dense)       (None, 1024)     4196352   
%		_________________________________________________
%		dropout (Dropout)     (None, 1024)     0         
%		_________________________________________________
%		dense_3 (Dense)       (None, 512)      1049088   
%		=================================================
%		Total params: 9,708,032
%		Trainable params: 9,708,032
%		Non-trainable params: 0
%		_________________________________________________
%		\end{verbatim}
%		\vspace{0.3cm}
%	\end{minipage}
%	
	\subsection{Training}
	
\begin{figure*}
		\includegraphics[width=0.5\textwidth]{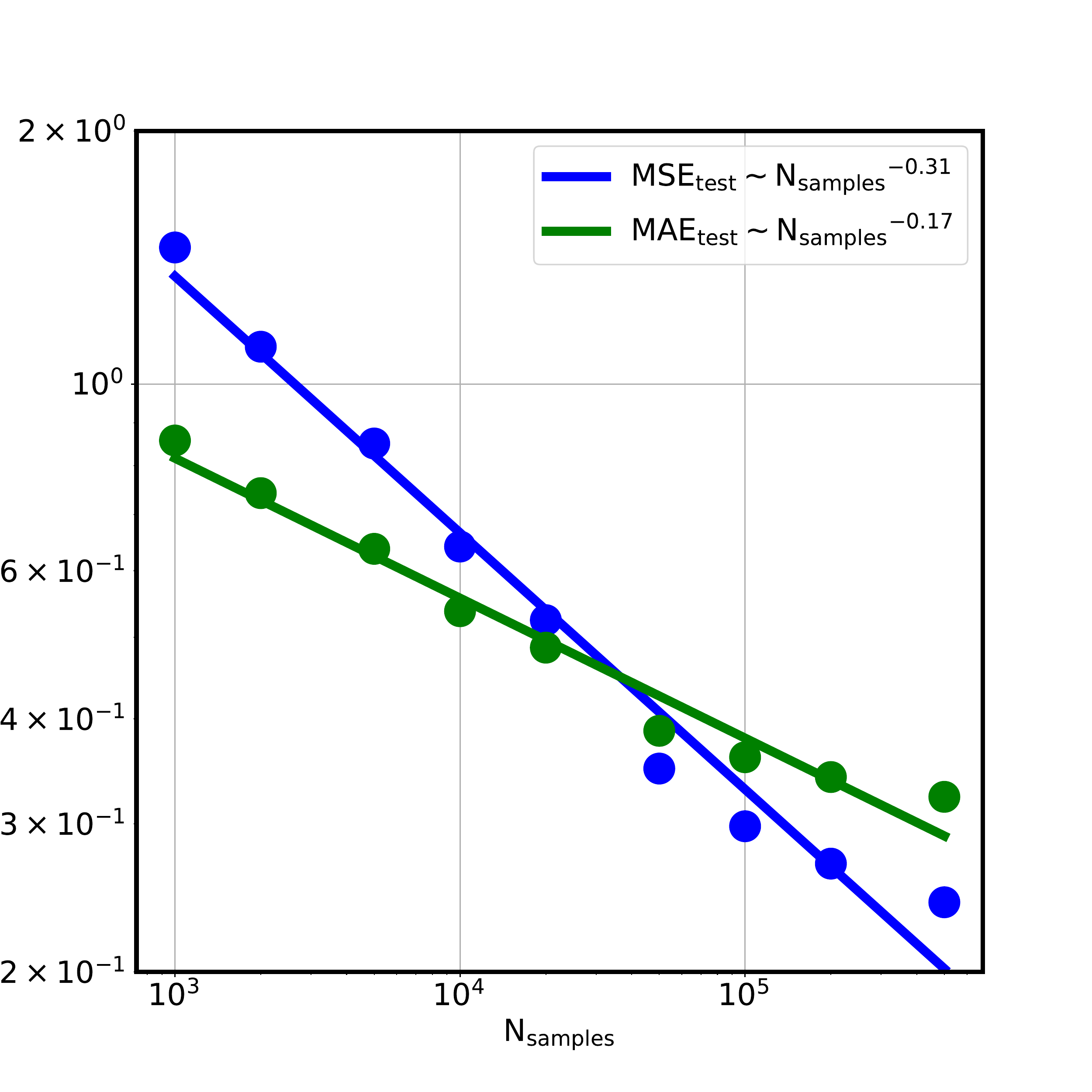}
		\includegraphics[width=0.5\textwidth]{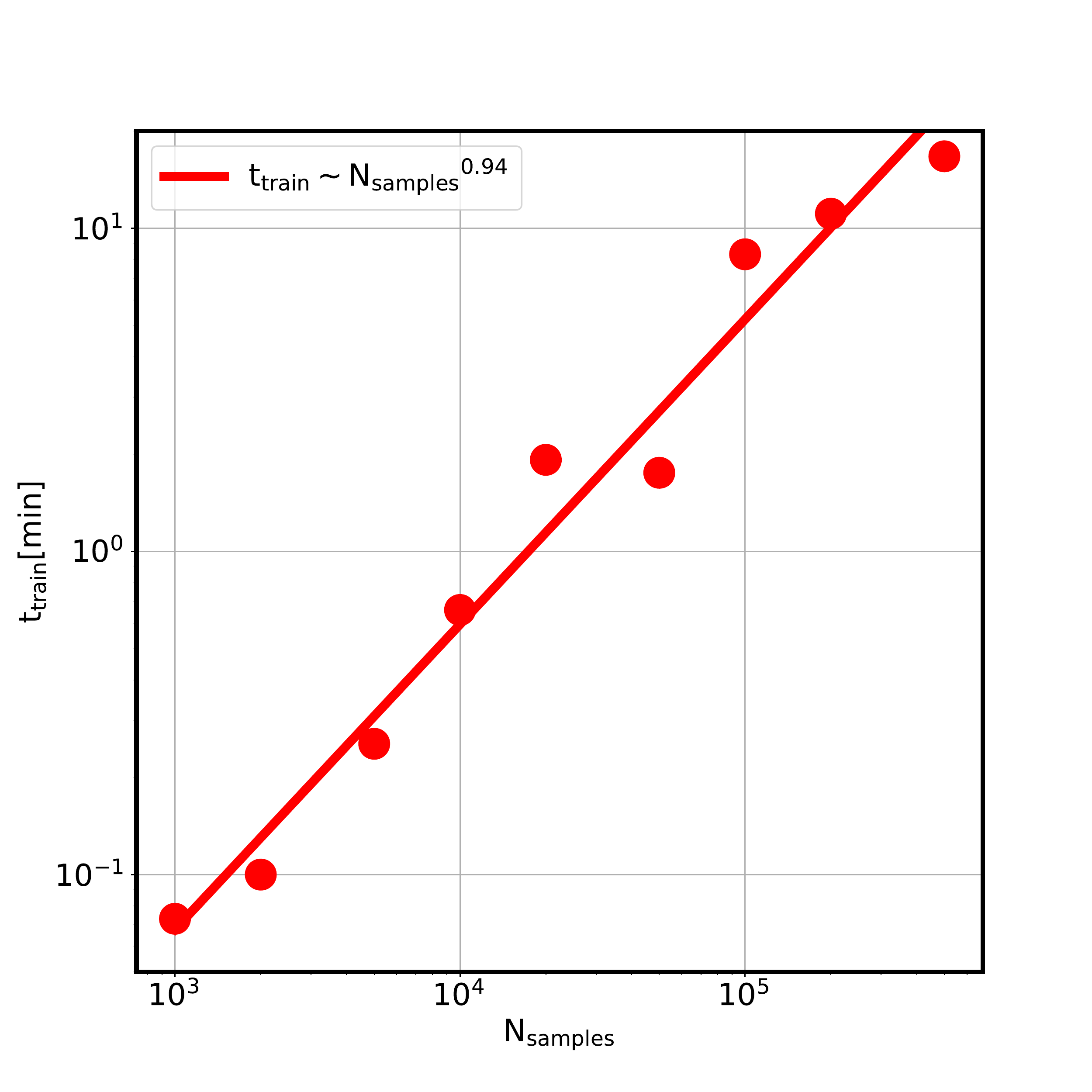}
		\caption{The MAE and the MSE are plotted against the number of training samples in the left panel. In the right panel, the plot shows the time it takes for a neural network to train, as function of the number of training samples. While the time it takes to train the ANN is directly proportional to the number of samples, the MAE and the MSE only decrease as weak power laws with exponents $\simeq -1/6$ and $-1/3$ respectively.}
		\label{fig: errors_and_times_wrt_nsamples}
	\end{figure*}

	At training time, the optimizer of choice is the Adaptive Moment Estimation algorithm (ADAM \cite{Kingma2014}), together with a mean-squared-error (MSE) loss function. All models are set to train for 500 epochs and an early stopping callback with a \texttt{patience} parameter value of 10 epochs is also provided. This callback prevents overfitting by stopping the ANNs training whenever the error in the validation set does not decrease anymore through the epochs. It is also important to define the number of samples needed to train: too few samples will not allow the network to learn; too many samples would waste time and resources. To this end, several preliminary experiments were carried out as well, in which the network training time, the MAE and the MSE were evaluated with respect to the number of training samples. Figure \ref{fig: errors_and_times_wrt_nsamples} shows that the time it takes to train the ANN is directly proportional to the number of samples, whereas the MAE and the MSE only decrease as weak power laws with exponents $-1/6$ and $-1/3$ respectively. This means that, to reduce the MSE and the MAE by some 10\% - 20\% at most, it would be required to double the sample size, which in turn would double the time to train and some juggling with the machine's memory would be required. All of this, assuming that the power law does not get weaker at larger number of samples. Weighing all these factors, a set of $N_s = 100K$ samples was used for training with each of the three artificial datasets used.

\section{Results}\label{sec: results}
	
	Three different datasets have been used to train three different neural networks: i) SLD profiles possessing odd symmetry, ii) single films and iii) lamellar structures. Thus, while all of the networks are architecturally equal, their learned weights are different.
	
	\subsection{Odd SLD profiles}\label{sec: odd_sld_profiles}
	
\begin{figure*}
		\includegraphics[width=1.0\textwidth]{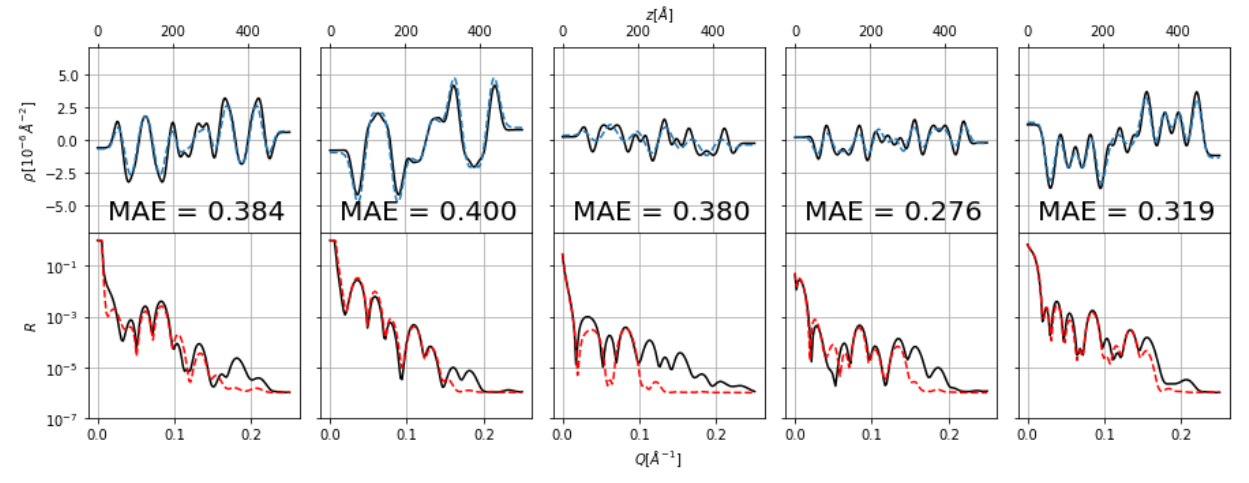}
		\caption{By training an ANN using a set of 100K random odd SLD profiles, the phase problem can be circumvented and there is a one to one correspondence between an SLD profile and its associated reflectivity curve. Five test reflectivity curves (black lines; bottom panels) were shown to the trained network for the first time; from those curves, the network guesses the SLD profiles that produce them (blue dashed lines; top panels). In the figure, the Mean Absolute Errors (MAE) with respect to the true SLD profiles (black lines; top panels) are also shown. Additional reflectivity curves are shown (red dashed lines, bottom panels), generated from the SLD profiles predicted by the ANN.}
		\label{fig: RandomOddAntisymmetricFunction100K}
	\end{figure*}

	If an SLD profile possesses the symmetry $\rho(z + L/2) = - \rho(z) + \rm{const}$, $z \in [0,L/2]$, the same reflectivity curve it produces is recovered by reflecting it at its mid-point (c.f. Section \ref{sec: phase_problem}). Thus, by defining a dataset composed only by odd SLD profiles, it is ensured that there is a one to one correspondence between an SLD profile and its associated reflectivity curve. Figure \ref{fig: RandomOddAntisymmetricFunction100K} shows five test SLD profiles recovered after a neural network is trained using such a dataset. The overall Mean Absolute Error for the test set, composed of 5K samples, lies at around 0.35.
	
	\subsection{Training on films with positive SLD}
	
	The simplest family of SLD profiles is that of single layers on top of a substrate. In this family, each SLD profile is made up of only three regions: The superstrate, $\rho_{sup} = 0$ extending from $z = - \infty$ to $z = \Delta_-$; the substrate, $\rho_{sub} = 1$ extending from $z = 512 - \Delta_+$ to $z = \infty$; and a film of constant SLD $\rho_{f}$ extending from $z = \Delta_-$ to $z = 512 - \Delta_+$. $\Delta_{+/-}$ are buffer regions defined to smoothen the transition between the SLD of the film and that of the super- and substrates. $\Delta_{+} = 50$ is fixed for all SLD profiles of this data set, while $\Delta_{-} \in [50,\Delta_{+})$ is randomly chosen for each generated profile and effectively defines the thickness of the film. The SLD of each film is randomly chosen $\rho_f \in [0,20]$.
	
\begin{figure*}
		\includegraphics[width=1.0\textwidth]{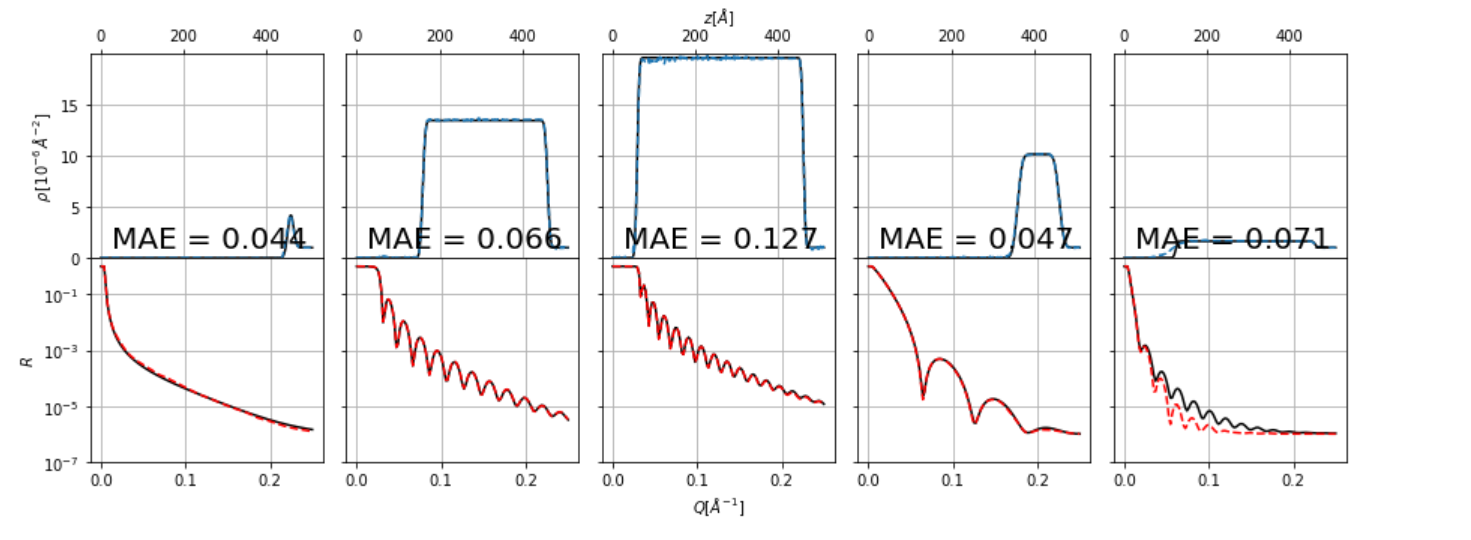}
		\caption{Single films are simple SLD profiles for which the trained network is able to correctly predict the associated SLD profiles.
			Five test reflectivity curves (black lines; bottom panels) were shown to the trained network for the first time; from those curves, the network guesses the SLD profiles that produce them (blue dashed lines; top panels). In the figure, the Mean Absolute Errors (MAE) with respect to the true SLD profiles (black lines; top panels) are also shown. Additional reflectivity curves are shown (red dashed lines, bottom panels), calculated from the SLD profiles predicted by the ANN.}
		\label{fig: RandomThinFilmFixedSubsAndSup100K}
	\end{figure*}

	Figure \ref{fig: RandomThinFilmFixedSubsAndSup100K} shows five test SLD profiles recovered after a neural network is trained using such a dataset. The overall Mean Absolute Error for the test set, composed of 5K samples, lies at around 0.08.
	
	\subsection{Training on Lamellar structures}\label{sec: lamellar results}
	
	The family of SLD profiles for this data set is defined by lamellar structures, each one having $n_r$ equally spaced regions of alternating SLDs between $\rho_1$ and $\rho_2$. Both SLD values were chosen randomly between -10 and 10, and the number of regions was randomly chosen between 1 and 64, extending from $z = \Delta$ to $z = 512 - \Delta$, where $\Delta = 50$ defines a buffer to smoothen the transition between the lamellar sample and its surroundings. The obtained profiles were smoothened using a Gaussian filter of a width randomly chosen between 0 and 10.
	
\begin{figure*}
		\includegraphics[width=1.0\textwidth]{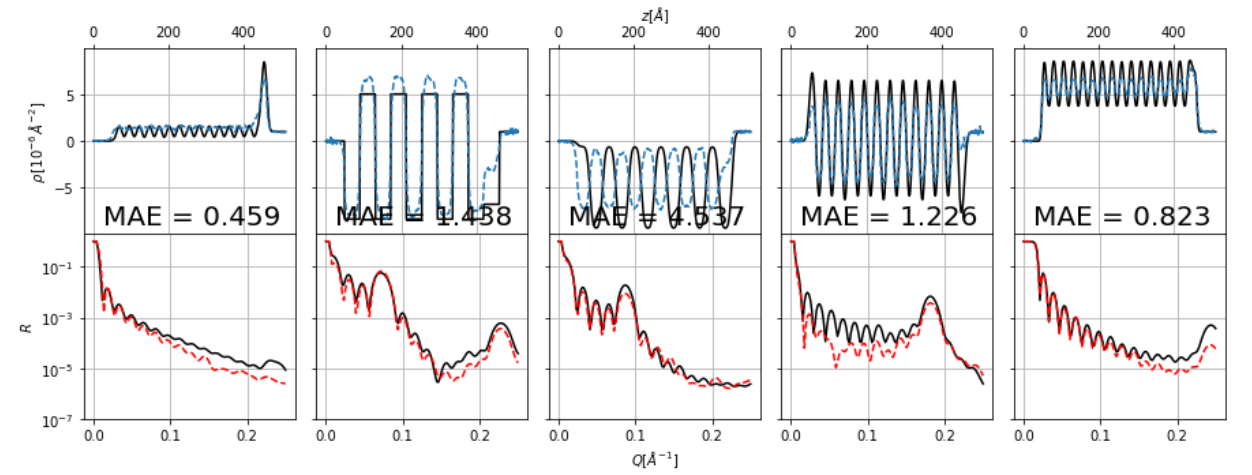}
		\caption{In the case of periodic lamellar SLD profiles, the phase problem is very present: even if most of the predictions recover to a close approximation the original reflectivity curves, the predicted SLD profiles differ from the actual targets.
			Five test reflectivity curves (black lines; bottom panels) were shown to the trained network for the first time; from those curves, the network guesses the SLD profiles that produce them (blue dashed lines; top panels). In the figure, the Mean Absolute Errors (MAE) with respect to the true SLD profiles (black lines; top panels) are also shown. Additional reflectivity curves are shown (red dashed lines, bottom panels), calculated from the SLD profiles predicted by the ANN.}
		\label{fig: RandomSmoothResonatorFixedSubsAndSup100K}
	\end{figure*}

	Figure \ref{fig: RandomSmoothResonatorFixedSubsAndSup100K} shows five test SLD profiles recovered after a neural network is trained using such a dataset. The overall Mean Absolute Error for the test set, composed of 5K samples, lies at around 0.98.
	
	The SLD profiles predicted by the ANN are not always consistent with the target SLD profiles. However, when calculating the reflectivity curves produced by such predicted profiles, the curve obtained is quite similar to the original curve presented to the ANN. For this family of SLD profiles, the trained network is a good example of a pseudo-inverse that, due to the degeneracy of the problem, is recovering a plausible solution but not necessarily the correct one.

	\subsection{Robustness against noise}\label{sec: robustness_noise}
	
	In order to train models robust against noise, the training set was \textit{augmented} by adding different levels of constant background, $\delta R$, modeled as Gaussian random white noise, to each of the ideally simulated reflectivity curves, $R_{\rm ideal}$. To avoid negative intensities, the noise was added to the square root of the reflectivity signal and the result squared to recover a noisy version of the original signal: 
	
	\begin{equation}\label{eq: adding_noise}
		R_{\rm noisy} = (\sqrt{R_{\rm ideal}} + X_{\mu, \sigma})^2,
	\end{equation}
	
	\noindent where $X_{\mu,\sigma}$ is a random variable extracted from a Gaussian distribution with $\mu = 0$ and $\sigma = \sqrt{\delta R}$.
	
\begin{figure*}
		\includegraphics[width=0.95\textwidth]{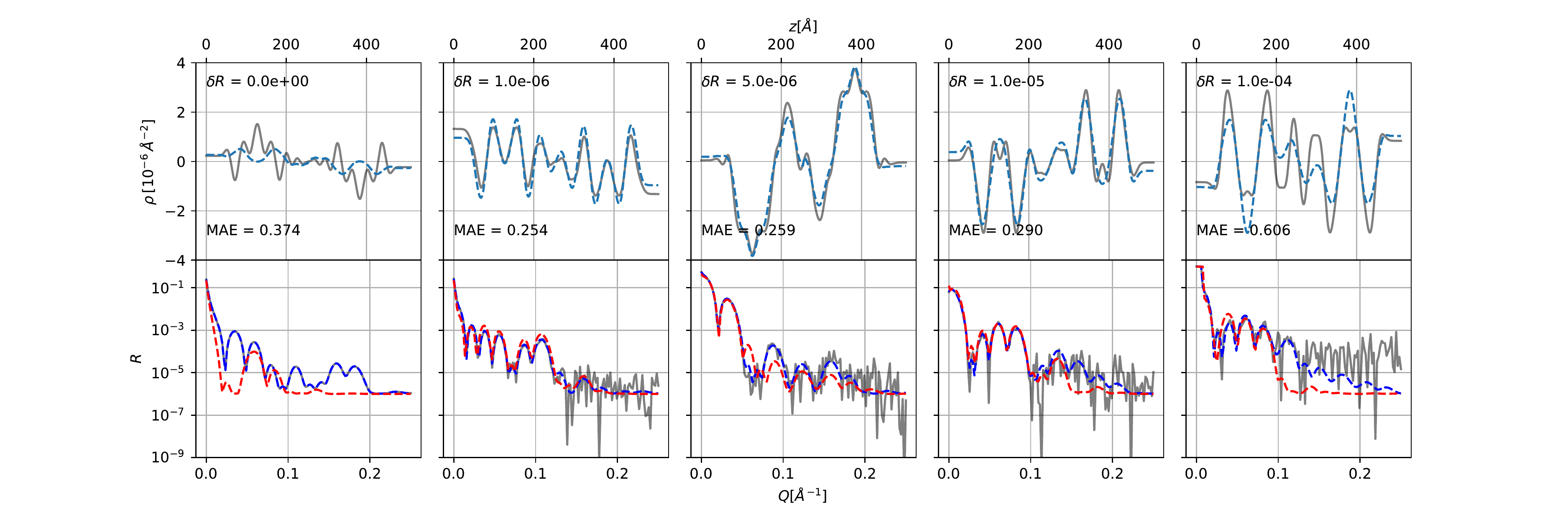}\\
		\includegraphics[width=0.95\textwidth]{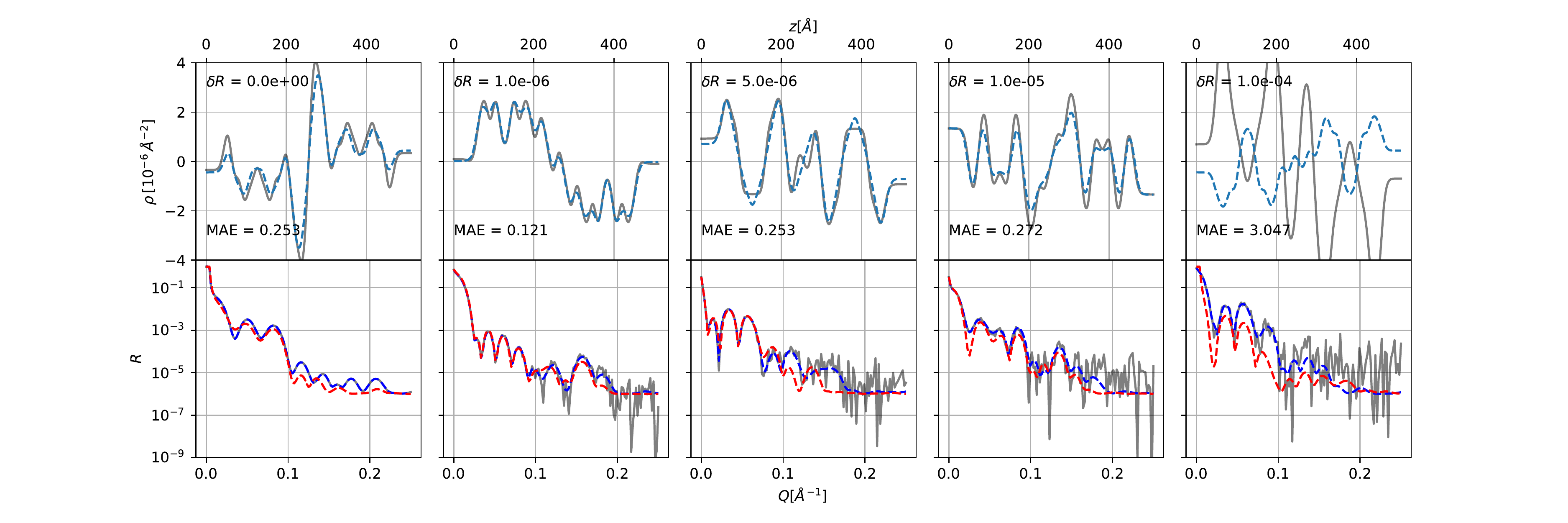}\\
		\includegraphics[width=0.95\textwidth]{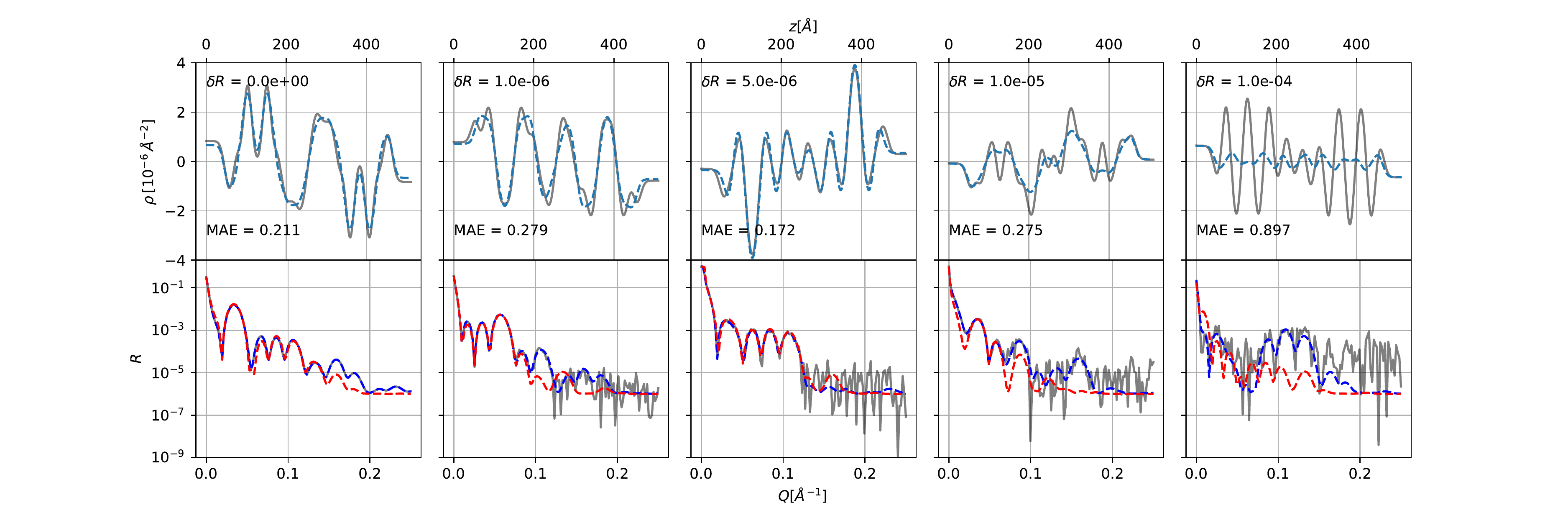}\\
		\includegraphics[width=0.95\textwidth]{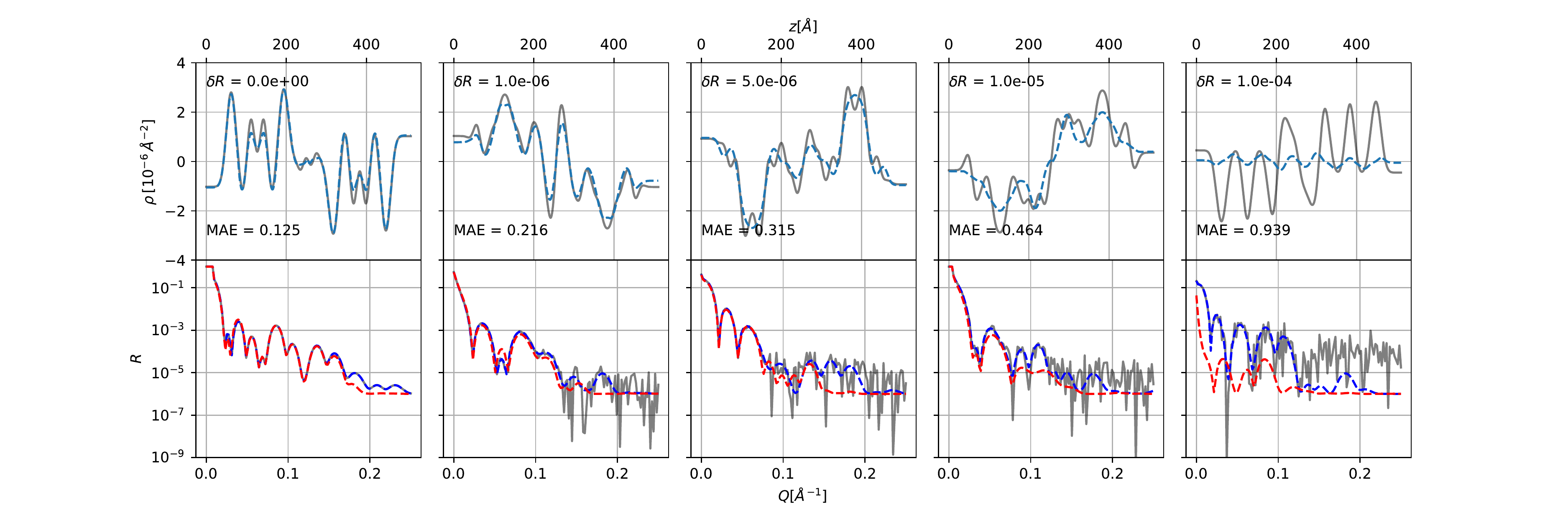}\\
		\vspace{-1cm}\caption{From top to bottom, four rows of panels are shown. In each panel, the top plot shows two SLD profiles and the bottom plot shows three reflectivity curves. The black reflectivity curves in the bottom plots mimic experimental data fed to the neural network, which predicts the blue SLD profiles in the top plots. The blue SLD profiles are compared to the target SLD profiles in black and the MAE is reported in each SLD plot. By simulating ideal and noiseless reflectivity curves, the target SLD profiles in black produce the blue reflectivity curves in the bottom plots, while the predicted SLD profiles in blue produce the red reflectivity curves. From left to right, the background noise intensity is $\delta R = \{0, 1, 5, 10, 100\} \times 10^{-6}$ respectively.}
		\label{fig: robustness_noise}
	\end{figure*}

\begin{figure}
		\includegraphics[width=0.5\textwidth]{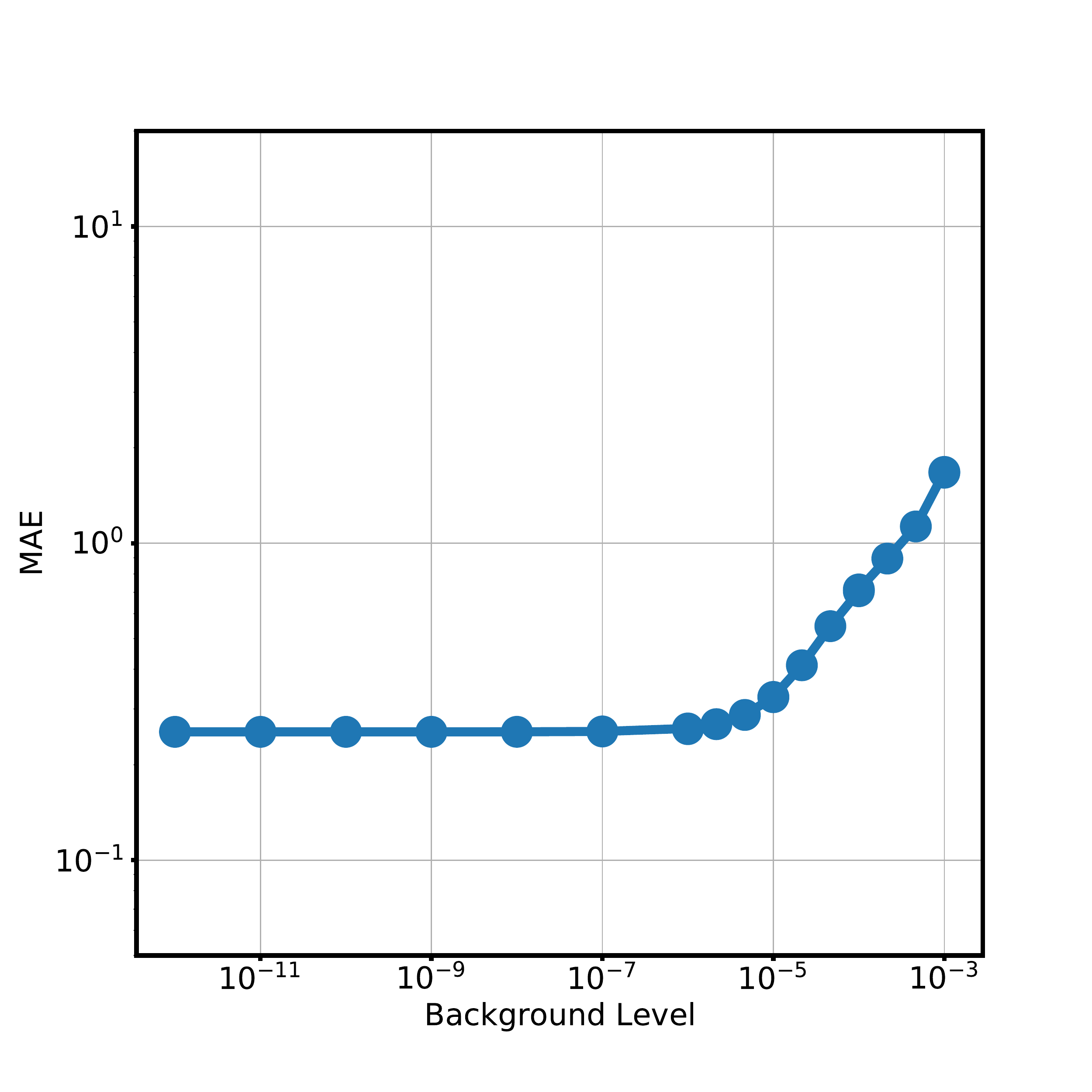}
		\caption{After being trained using a noise-augmented dataset, the neural network is able to recover SLD profiles from noisy reflectivity curves with a Mean Absolute Error of around 0.3, provided that the noise present in the reflectivity curves fed to the network satisfies $\delta R \lesssim 10^{-5}$. For increasing noise levels, the error incurred by the network steeply rises, so its predictions are no more reliable.}
		\label{fig: mae_vs_noise}
	\end{figure}

	The training was carried out over the augmented data set of artificial SLD profiles of Section \ref{sec: odd_sld_profiles}. The dataset consisted of 10K clean reflectivity curves, replicated 9 times for each level of background $\delta R \in \{1,\, 2,\, 3,\, 4,\, 5,\, 10,\, 20,\, 50,\, 100\} \times 10^{-6} $, for a total of 100K reflectivity curves. After training, the model was tested over never-seen before reflectivity curves with arbitrary levels of noise, $10^{-6} < \delta R < 10^{-3}$, delivering satisfactory results for background noise of up to $\delta R \simeq 10^{-5}$, and performing poorly for reflectivity signals with stronger background components, as can be observed in Figure \ref{fig: robustness_noise} and Figure \ref{fig: mae_vs_noise}.
	
	\subsection{A parameter estimation for a two-layer system}
	
\begin{figure}
		\includegraphics[width=0.5\textwidth]{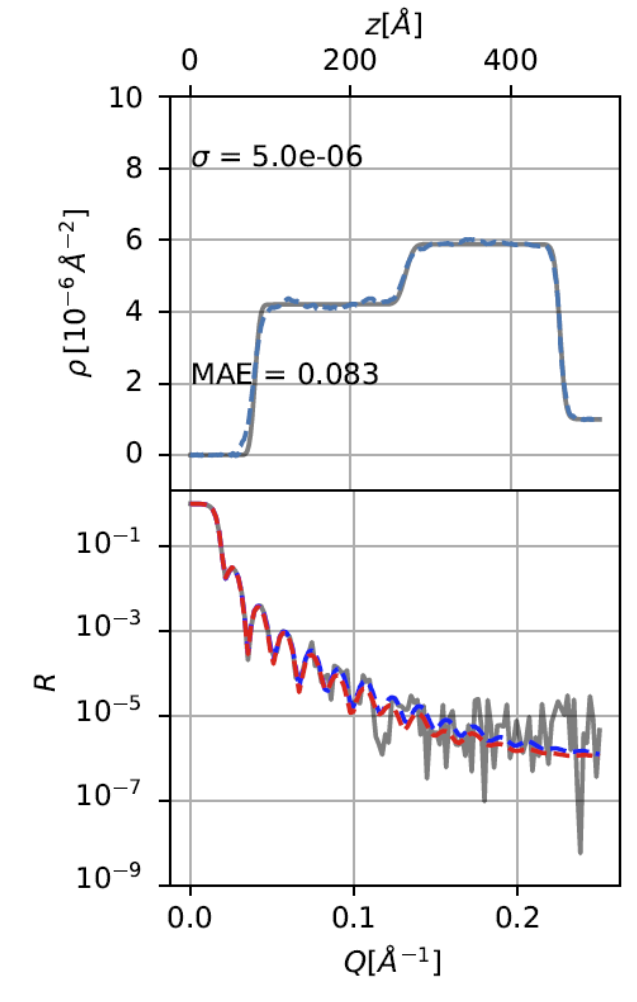}
		\caption{In a real experiment, the neural network would be fed with the experimental curve (in black, bottom plot) and produce a guess SLD profile (in blue, top plot). Using a simulation package, a reflectivity curve simulated from the guess SLD profile (in red, bottom plot) could then be compared to the experimental data. The ideal SLD profile (in black, top plot) and the noisless reflectivity curve it produces (in blue, bottom plot) are also shown for completeness, together with the noise level of the fake reflectivity curve and the MAE the neural network produced while guessing the correct SLD profile.}
		\label{fig: simulated_experiment_bilayer}
	\end{figure}

	In a real parameter estimation workflow, it is necessary to provide a quantitative analysis including the goodness of fit for several parameters (e.g. the thickness of each layer, the SLD value of each layer and the interfacial roughness between each couple of layers). Available software providing such a functionality requires that, for each sample, a model is built by specifying all the parameters to fit. In contrast, the proposed approach is, in this sense, parameter free. A drawback of the proposed approach is then its lack of a quantitative parameter estimation machinery. However, it offers the advantage of its immediate response: thousands of reflectivity curves can be inverted to SLD profiles in a couple of seconds. As such, the proposed approach should be taken not as a replacement for available fitting methods, but rather as a preliminary tool to accelerate the process, providing preliminary glimpses of what the ultimate parameters of a traditional fitting process may look like. Figure \ref{fig: simulated_experiment_bilayer} shows a simulated reflectivity curve (a fake experimental curve) which, after being fed to the neural network produces an SLD profile from which the experimenter can infer the following parameters: 2 layers of thicknesses of around $200 \AA$ and $150\AA$ each, with corresponding SLD values of around $\rho = \{4,6\} \times 10^{-6} \AA^{-2}$; a substrate SLD of $\rho_{\rm sub} = 1\times 10^{-6} \AA^{-2}$ and a superstrate of SLD $\rho_{\rm sup} = 0$. These rough estimations can be then fed into a traditional model minimizer as first guesses to the fitting process. 
	
	\section{Discussion}\label{sec: discussion}
	
	The time required to generate each of the training data sets and to train each of the ANNs is rather short: around a couple of hours for data generation and a similar time for ANN training (provided a GPU is available \footnote{In the present work, an Nvidia GeForce GTX 1080 was used for the production runs, while several prototypes were developed on Google Cloud infrastructure.}). For unfortunate cases in which the data set is not suited for ANN training (because of the phase problem, for example), the overfitting regime may be reached rather soon and the training stops even in less than 10 minutes.  
	
	Typical trained ANNs require only around 100 MB of disk space, whereas the training data is at least around 500 MB large (of reflectivity curves alone). Once trained, the ANNs are able to recover plausible SLD profiles from 5K reflectivity curves in around 0.5 sec. What these metrics imply, is that the ANNs are abstracting in an efficient way the information contained in the training data: as if they were compressing the whole dataset (and more) with a compression ratio of at least 5:1. 
	
	A good question to ask now is whether the ANNs are actually learning the transformations or are only memorizing the associations between particular SLD profiles and reflectivity curves. While the evolution of the metrics during the training and the performance on the test set suggest that actual learning is taking place (See \ref{sec: metrix_evolution}), it must be stressed that the ANNs are certainly not learning a full general transformation, but are rather optimized for interpolating within the family of SLD profiles for which they are trained. Thus, in order to take advantage of the proposed approach in real experiments, the simulated training data should incorporate the instrument specifications with as much detail as possible. Additionally, in order to train an ANN robust against noise, for each target SLD profile, several noisy reflectivity curves with varying levels of noise were simulated, effectively allowing the network to give higher priority to data points lying at lower $Q$ values. In Section \ref{sec: robustness_noise}, it is shown that such an approach to neural network training succeeds in allowing the network to recover SLD profiles provided that the background noise satisfies $\delta R \lesssim 10^{-5}$ --a constraint usually fulfilled. The trends reported, in fact, demonstrate the possibility of training noise robust networks.
	
	Due to the quick response of the trained ANNs, they could be incorporated in a data pipeline able to potentially provide preliminary analyses of real time phenomena like the swelling or drying of thin films (See also \cite{Greco2019}), or used in large batch analysis of huge data sets coming from large facilities. In fact, the time required for an ANN to invert 10K reflectivity curves is only around one second. However, as seen in Section \ref{sec: lamellar results}, interpreting reflectivity curves in such a way, may offer plausible SLD profiles that nevertheless do not correspond to the actual sample under the beam, thus, a careful examination at a later stage will be always required and, more likely than not, the ANN-inferred SLD profiles should be used for the time being as starting models to feed traditional fitting methods. In the future, with a higher degree of customization, ANNs could be incorporated into an automatized analysis pipeline. 
	
\begin{figure*}
		\includegraphics[width=1.0\textwidth]{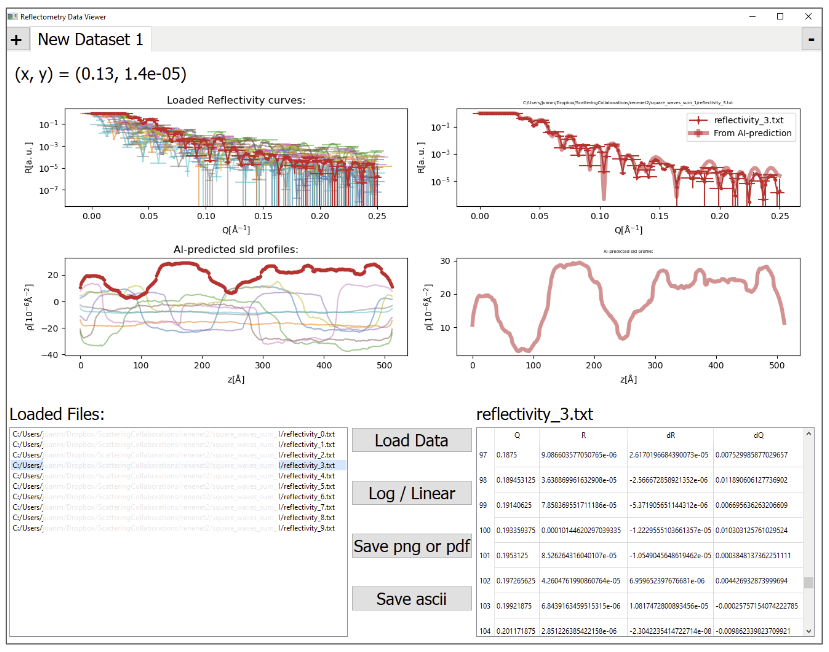}
		\caption{In a real data analysis workflow, several reflectivity curves (top left plot) are loaded from several files (bottom left list), and the neural network is automatically engaged to produce possible SLD profiles (bottom left plot), from which the experimenter can \textit{eyeball-estimate} the parameters to feed traditional minimizers. The contents of the selected file are also shown (bottom right table), together with the isolated inferred SLD profile (bottom right plot) and two isolated reflectivity curves (top right plot) corresponding to the data and the simulated from the ANN inferred SLD profile. See also \url{https://youtu.be/NKismnONlmA}}
		\label{fig: data_loader_prototype}
	\end{figure*}

	The proposed algorithm has the advantage of being quick for batch analyses at the cost of losing a high degree of per-profile customization. In fact, the quick response of neural networks, makes it possible for the inferred SLD profiles to be readily shown to experimenters at data-loading time, together with the loaded reflectivity curves. In this way, by being offered with plausible SLD profiles, experimenters could easily infer from them the relevant parameters with which to feed traditional fitting models. Figure \ref{fig: data_loader_prototype} offers a sketch of what such a data loader could look like: a small piece of software that could be included as an optional add-in in some of the available software packages for reflectometry data analysis (e.g. Bornagain and Refnx \cite{BornAgain2020, Refnx2019}, among others).
	
	\section{Conclusions}

	In the present work, sample physical models are described under a new paradigm: detailed layer-by-layer quantitative descriptions (SLDs, thicknesses, roughnesses) are replaced by parameter free curves $\rho(z)$, allowing a-priori assumptions to be fed in terms of the sample family to which a given sample belongs (e.g. ``thin film'', ``lamellar structure'', etc.). Such an approach is not comparable to traditional fitting methods and acts as a preliminary analysis tool available for experimenters to make quick first estimates of the relevant parameters to feed traditional fitting methods.
	
	Though the ANNs described in this work are not yet capable of interpreting real reflectivity curves in terms of SLD profiles, the numerical experiments carried out show that the short training times, the small space required and the ANN robustness against noise, could make it practical to train several ANNs with richer architectures, specialized in different instruments and selected families of samples. In order to bring the proposed approach to the arena of real experiments, additional information about the instrument such as resolution and $Q$ range would also be needed.
	
	\ack
	
	JMCL would like to thank the BornAgain team at JCNS-MLZ, especially Joachim Wuttke for his healthy skepticism and criticism; and Alexander Schober for his optimistic encouragement. I would also like to thank, for the many fruitful discussions, Wojciech Potrzebowski from the ESS; Miguel González from the ILL; Alexandros Koutsioumpas, Jean F. Moulin, Gaetano Mangiapia, and Martin Haese from the MLZ. Their feedback and comments have certainly had a positive impact on the development and presentation of this work.
	
	\section*{Data Availability Statement}
	Data sharing is not applicable to this article as no new data were created or analysed in this study.
	
	\hfill\\
	{\textbf{References}}
	
	\bibliographystyle{iopart-num}
	\bibliography{mybib}

	\appendix
	\section{Appendix}
	
	\subsection{Training on random SLD profiles}
	
	\begin{figure*}
		\includegraphics[width=1.0\textwidth]{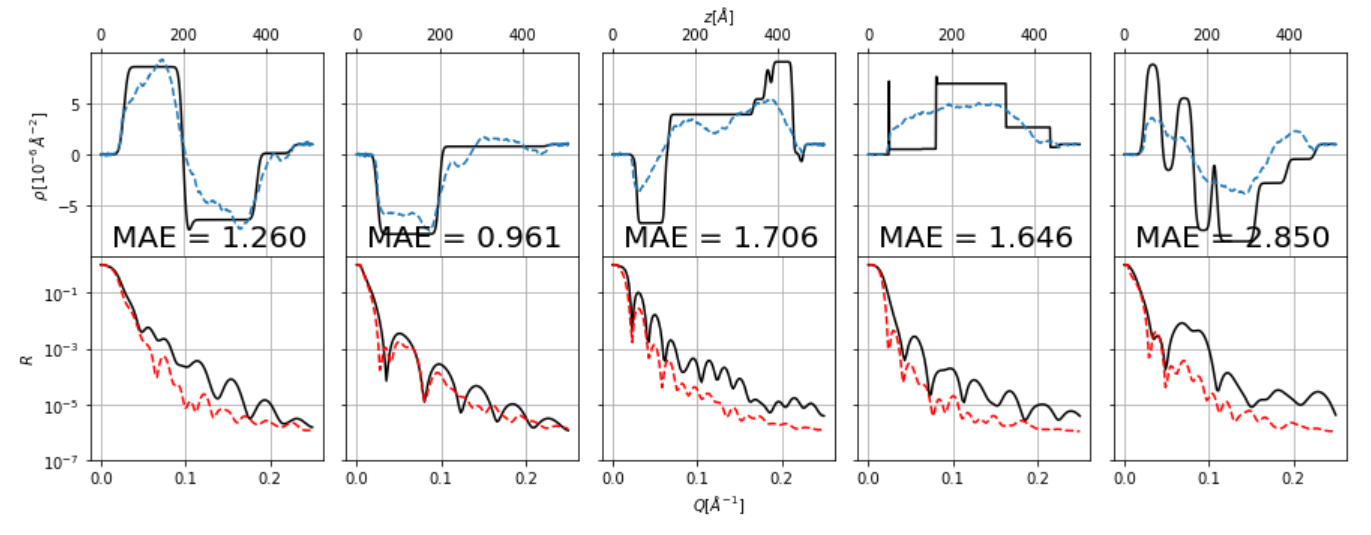}
		\caption{The training of the ANN fails when using a set of 100K random SLD profiles of between 1 and 16 slabs of random thicknesses and heights between -10 and 10. Five test reflectivity curves (black lines; bottom panels) were shown to the trained network for the first time; from those curves, the network guesses the SLD profiles that produce them (blue dashed lines; top panels). In the figure, the Mean Absolute Errors (MAE) with respect to the true SLD profiles (black lines; top panels) are also shown. Additional reflectivity curves are shown (red dashed lines, bottom panels), calculated from the SLD profiles predicted by the ANN.}
		\label{fig: RandomSmoothSlabsFixedSubsAndSup100K}
	\end{figure*}
	
	Throughout this work, we have argued that the ANN training process is sure to fail in a majority of cases, as the solution space usually has branches and many of the training targets may lie in inconsistent branches of the solution space, causing the weights of the ANN to drift in inconsistent directions during the training. To prove this point, an additional ANN was trained using a set of SLD profiles composed of a random number of layers of random thicknesses and random heights. Figure \ref{fig: RandomSmoothSlabsFixedSubsAndSup100K} shows five test SLD profiles recovered after a neural network is trained using such a dataset. None of the predicted SLD profiles is consistent with the targets and none of the recovered reflectivity curves is consistent with the input data. The Mean Absolute Error for the test set, composed of 5K samples, lies at around 2.4.
	
	\subsection{Metrics Evolution}\label{sec: metrix_evolution}
	
	%\textit{
	After the ANN architecture is defined, the training proceeds, using the training set to modify the weights of the network and the validation set to asses its performance on unseen data to avoid over-fitting. In this work, the metric used to keep track of the performance of the ANN is the Mean Absolute Error (MAE), which is computed after each epoch.
	%}
	
\begin{figure*}
		\includegraphics[width=0.5\textwidth]{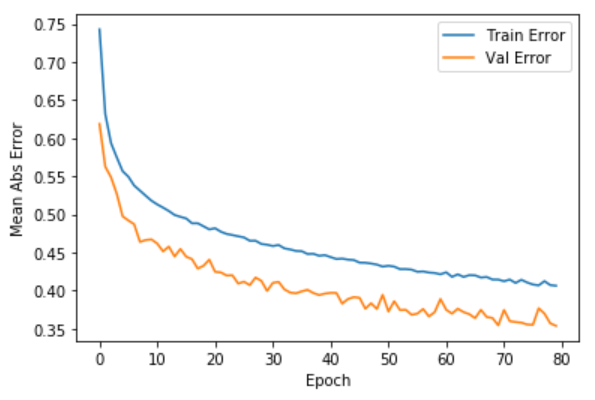}
		\includegraphics[width=0.5\textwidth]{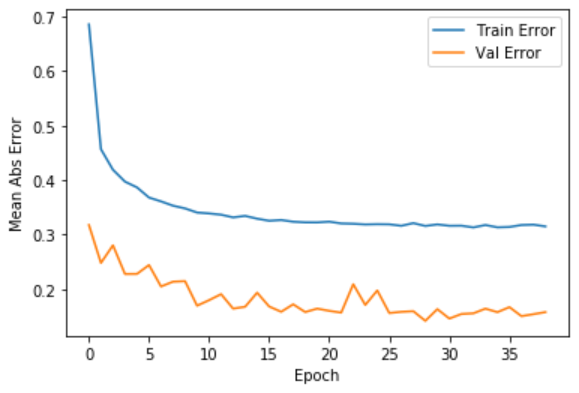}
		\includegraphics[width=0.5\textwidth]{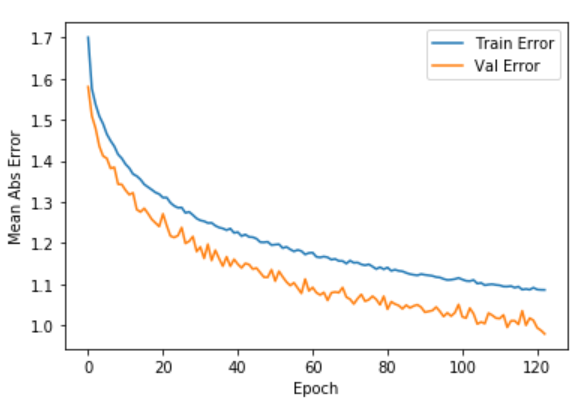}
		\includegraphics[width=0.5\textwidth]{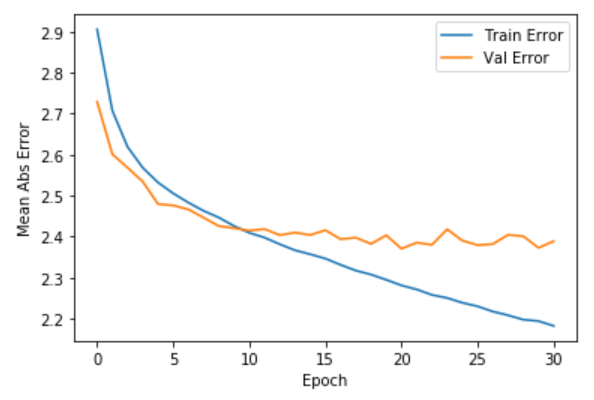}
		\caption{The MAE evolution throughout the epochs as the training proceeds is shown for the different families of SLD profiles studied in this work: Odd functions (top left panel), single films (top right panel), lamellar structures (bottom left panel) and random SLD profiles (bottom right panel). See the text for discussion.}
		\label{fig: MetricsEvolution}
	\end{figure*}

	%\textit{
	The smoking gun of an over-fitting network is a higher performance (lower error) reported for the training set than that reported for the validation set. A healthy network, instead, performs as well for the validation set as it does for the training set (in fact, the performance during training is usually a bit better than during validation), and the curves that follow both errors evolve in a similar fashion as epochs progress. To assess the health of the training processes carried out in the present work, the evolution of the MAE for each of the datasets is analyzed in this section (See Figure \ref{fig: MetricsEvolution}).
	%}
	
	%\textit{
	In the present work, there is one case of an over-fitting network, namely, the network trained on pure random data: its performance continues to improve at every epoch for the training set, while stalling at a constant error for the validations set. This behavior demonstrates that the network is memorizing the training set, rather than learning a pseudo-inverse transformation. As discussed in Section \ref{sec: data simulation}, this is to be expected since, due to the ill-posedness of the problem, a pseudo-inverse simply does not exist in this case.
	%}
	
\begin{figure}
		\includegraphics[width=0.5\textwidth]{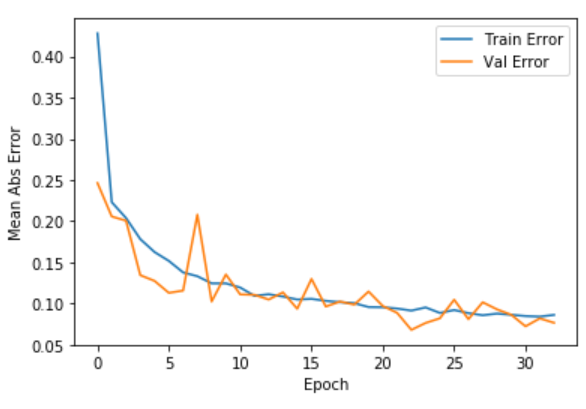}
		\caption{For the case of single films SLD profiles, the MAE evolution throughout the epochs as the training proceeds is shown for a special ANN in which the dropout layer is not present. See the text for discussion.}
		\label{fig: MetricsEvolutionNoDropout}
	\end{figure}

	%\textit{
	The errors reported for the rest of the networks show a lower performance (higher error) for the training set than that shown for the validation set. Such a behavior implies that the network is performing better on unknown data (validation set) than on known data (training set), which is counterintuitive. The hypothesis is that such a counterintuitive behavior comes from the dropout layer which, being activated only at training time, lowers the performance on the training set --a condition that is rather perceived as better validation performance. To test the hypothesis, an extra experiment was carried out in which the dropout layer was removed from an ANN and later trained over the single films data set. The evolution of the error during the training process proceeded as that of a healthy network described above and the expected relation between training and validation errors was recovered: $\mathrm{MAE_{valid}} \leq \mathrm{MAE_{train}}$ (See Figure \ref{fig: MetricsEvolutionNoDropout}).
	%}
	
\end{document}